# ORBITAL PERIOD CHANGES IN WZ SAGITTAE

Joseph Patterson,[1] Geoffrey Stone,[2] Jonathan Kemp,[3,4] David Skillman,[5]
Enrique de Miguel,[6,7] Michael Potter,[8] Donn Starkey,[9] Helena Uthas,[10] Jim Jones,[11]
Douglas Slauson,[12] Robert Koff,[13] Gordon Myers,[14] Kenneth Menzies,[15] Tut Campbell,[16]
George Roberts,[17] Jerry Foote,[18] Tonny Vanmunster[19], Lewis M. Cook,[20] Thomas Krajci[21],
Yenal Ogmen,[22] Richard Sabo[23], & Jim Seargeant[24]

[1] Department of Astronomy, Columbia University, 550 West 120th Street, New York, NY 10027; jop@astro.columbia.edu
[2] CBA–Sierras, 44325 Alder Heights Road, Auberry, CA 93602; geofstone@earthlink.net
[3] Mittelman Observatory, Middlebury College, Middlebury, VT 05753; jkemp@middlebury.edu
[4] Visiting Astronomer, Cerro Tololo Inter-American Observatory, National Optical Astronomy Observatories, which is operated by the Association of Universities for Research in Astronomy, Inc., (AURA) under cooperative agreement with the National Science Foundation
[5] CBA–East, 159 Research Road, Greenbelt, MD 20770; dskillman@comcast.net
[6] Departamento de Física Aplicada, Facultad de Ciencias Experimentales, Universidad de Huelva, 21071 Huelva, Spain
[7] CBA–Huelva, Observatorio del CIECEM, Parque Dunar, Matalascañas, 21760 Almonte, Huelva, Spain; edmiguel63@gmail.com
[8] CBA–Baltimore, 3206 Overland Avenue, Baltimore, MD 21214; mike@beverlyhillsobservatory.org
[9] CBA–Indiana, DeKalb Observatory (H63), 2507 County Road 60, Auburn, IN 46706; donn@starkey.ws
[10] Viktor Rydberg–Djursholm, Viktor Rydbergs väg 31, 182 62 Djursholm, Sweden; helena.uthas@vrg.se
[11] CBA–Oregon, Jack Jones Observatory, 22665 Bents Road NE, Aurora, OR 97002; nt7t@centurytel.net
[12] CBA–Iowa, Owl Ridge Observatory, 73 Summit Avenue NE, Swisher, IA 52338; dmslauson@netscape.net
[13] CBA–Colorado, Antelope Hills Observatory, 980 Antelope Drive West, Bennett, CO 80102; bob@antelopehillsobservatory.org
[14] CBA–San Mateo, 5 Inverness Way, Hillsborough, CA 94010; gordonmyers@hotmail.com
[15] CBA–Framingham, 318A Potter Road, Framingham, MA 01701; kenmenstar@gmail.com
[16] CBA–Arkansas, 7021 Whispering Pine, Harrison, AR 72601; jmontecamp@yahoo.com
[17] CBA–Tennessee, 2007 Cedarmont Drive, Franklin, TN 37067; georgeroberts0804@att.net
[18] CBA–Utah, 4175 East Red Cliffs Drive, Kanab, UT 84741; jfoote@scopecraft.com
[19] CBA–Belgium, Walhostraat 1A, B-3401 Landen, Belgium; tonny.vanmunster@gmail.com
[20] CBA–Concord, 1730 Helix Court, Concord, CA 94518; lew.cook@gmail.com
[21] CBA–New Mexico, PO Box 1351 Cloudcroft, NM 88317; tom_krajci@tularosa.net
[22] CBA–Cyprus, Green Island Observatory (B34), Geçitkale, North Cyprus; yenalogmen@yahoo.com
[23] CBA–Montana, 2336 Trailcrest Drive, Bozeman, MT 59718; richard@theglobal.net
[24] CBA–Edgewood, 11 Hilltop Road, Edgewood, NM 87015; jimsarge@gmail.com

**ABSTRACT**

We report a long-term (1961–2017) study of the eclipse times in the dwarf nova WZ Sagittae, in an effort to learn its rate of orbital-period change. Some wiggles with a time scale of 20–50 years are apparent, and a connection with the 23-year interval between dwarf-nova eruptions is possible. These back-and-forth wiggles dominate the O–C diagram, and prevent a secure measurement of the steady rate of orbital-period change.

> The line, it is drawn, the curse, it is cast.
> The slow one now will later be fast...
> For the times, they are a-changin'.
> — Dylan (1963)

## 1 INTRODUCTION

WZ Sagittae is the world's most famous cataclysmic variable. This one star has been the sole subject of over 130 papers in the refereed journals, plus a few hundred unrefereed (conferences, abstracts, observing reports and proposals). The earliest detailed studies (Greenstein 1957, Krzeminski & Kraft 1964, Krzeminski & Smak 1971, Faulkner 1971) established its basic structure: a dwarf nova with a very short binary period (81 minutes) and a very long interval between outbursts (~30 years).

There are many reasons for the star's allure. It is the nearest dwarf nova ($d$ = 43 pc, Thorstensen 2003), and therefore the brightest ($V$=7.5 in outburst). It was the first cataclysmic variable (CV) in which the "blue star" was recognized to be a white dwarf (Greenstein 1957). Its accretion disk displays a multitude of rapidly moving emission lines, which, together with the eclipses, enable good constraints on the binary motions (Krzeminski & Smak 1971, Steeghs et al. 2007). It flashes a complex spectrum of periodic signals in outburst; these are the so-called "superhumps", signifying precession of the accretion disk (Patterson et al. 1981, 2002). It's the prototype of the most common type of cataclysmic variable — the "WZ Sge stars", comprising ~70% of all CVs (Patterson 1998, 2011). The donor star in the binary appears to be of very low mass, probably in the range 0.06–0.09 $M_\odot$. And many studies consider it to be one of the oldest of all CVs, with mass transfer powered by the bare minimum — through angular momentum loss by gravitational radiation (GR). Recent detailed reviews of WZ Sge and its class are given by Kuulkers et al. (2011) and Kato (2015).

In short: it's the *Andromeda Galaxy* of CVs.

Han et al. (2017a, hereafter H17) studied the eclipse timings over 55 years, and concluded that the star is decreasing its orbital period, at a rate similar to that predicted by the assumption that GR powers the mass transfer. This would be an important and somewhat surprising result, because WZ Sge is often considered a prototype "period bouncer" — in which the GR-powered mass transfer causes the orbital period to increase. So we undertook to evaluate this, by studying our large archive of time-series photometry.

## 2 THE OBSERVATIONS

We have been acquiring time-series photometry of many CVs over the last ~30 years, mainly with the globally-distributed telescopes of the Center for Backyard Astrophysics (CBA). Despite small apertures (mostly 0.2–0.4 m) and long integration times (usually ~30 s), these telescopes are well-suited for study of CVs as faint as $V$~17.5. This is partly because our campaigns are so extensive — usually covering one star for many years — and partly because we typically observe in white light (4500–8500 Å, hence more accurately "pink"). This yields 6–8 times more photons than the typical "wide-band" filters more commonly employed in astronomical photometry. The worldwide distribution of telescopes also immunizes us from problems in the daily aliasing of periodic signals. Further details on CBA data collection and analysis are given by Skillman & Patterson (1993) and de Miguel et al. (2016).

At quiescence, the light curve displays shallow but sharp eclipses on the 81-minute binary period. These eclipses may well be a suitable and stable marker of binary phase, because they are present in all the historical studies, as well as the ~3000 orbits covered in our studies (Patterson 1980; Patterson et al. 1981, 1998, 2002; and the present paper). The lower frame of Figure 1 shows a typical spliced nightly light curve, and the upper frame shows the average orbital light curve during 2017, at 50-s time resolution. The various light curves are normalized to "zero" magnitude out of eclipse; this corresponds roughly to $V$=15.5, but varies slightly from night to night. Also, the calibration to a standard $V$ is uncertain by ~0.2 mag (because of the very wide bandpass). These light curves are typical of those in the historical record.

With the small-telescope data, we compute averages over ~10–15 orbits; this is usually necessary to enable an accurate measurement of mid-eclipse. Figure 2 shows a typical 15-orbit average, at 5-second effective time resolution. Despite the small apertures and coarseness of the original time resolution, this captures the major features of the eclipse (shape and timing; see the Figure 2 caption for more discussion). Some time-series data was also obtained with larger telescopes (the 1.3 m and 2.4 m telescopes of MDM Observatory, and other telescopes at Lick Observatory and Cerro Tololo Inter-American Observatory). That data has time resolutions in the range 1-5 s, and we derive eclipse timings by averaging over 2–3 orbits.

Following Robinson et al. (1978, hereafter RNP), we define the time of mid-eclipse as the average of mid-ingress and mid-egress times. This enables us to take advantage of the relatively sharp walls of the eclipse, and is the procedure used by all previous studies. However, timings during July–November 2001 showed very large residuals, presumably because the star had not yet reached quiescence after its 2001 eruption. That post-eruption data has been extensively discussed (Patterson et al. 2002), but is irrelevant here, and therefore excluded.

## 3  ANALYSIS FOR PERIOD CHANGE

In principle, WZ Sge offers a quite sensitive test for orbital-period change. Eclipse timings now span 56 years (1961–2017), and most show an rms dispersion of less than ~10 s in all seasons of good coverage. Thus the timings are significantly affected by leap seconds, and somewhat also by barycentric correction. So the relevant uniform time scale is BJD(TDB), and we have converted all times to that scale, using the calculator of Eastman et al. (2010). The earliest eclipse times (Krzeminski & Kraft 1964, Krzeminski & Smak 1971) have greater scatter, so we have consolidated nearby — usually three consecutive or nearly so — timings in that data to produce 20, which show a scatter similar to that of the later timings. Most of the later timings (1971–2017) are taken at face value, except for those listed in Table 2 of Patterson et al. (1998).

We have re-measured the latter timings to improve accuracy, and averaged over consecutive eclipses where the noise level was somewhat higher.

The whole collection is listed in Table 1. Figure 3 is an O–C diagram of the timings with respect to a test ephemeris, which is

$$\text{Mid-eclipse} = \text{BJD(TDB)}\ 2{,}437{,}547.72863 + 0.056687847\ E. \qquad (1)$$

By convention, we take the first observed eclipse time to define the epoch (starting point of the ephemeris).

It is impossible to estimate the timing uncertainty for each eclipse. The eclipses are shallow (0.15 mag deep) and often asymmetric, and are afflicted by random flickering of 0.05–0.10 mag. The RNP data illustrate the eclipse profiles, and are the best and most extensive published data of high temporal resolution. The scatter of their timings with respect to Eq. (1) is shown in Figure 4, where the two seasons of dense coverage are separately labeled. Each season shows an internal scatter of only 4 s, but the two seasons differ by 12±3 s (= 0.0025 cycles). So this could be considered as an estimate of potential "seasonal noise" in the timings.

From the 167 eclipse timings in Table 1, the resultant linear ephemeris is

$$\text{Mid-eclipse} = \text{BJD(TDB)}\ 2{,}437{,}547.72891(4) + 0.05668784689(4)\ E, \qquad (2)$$

and is the straight line in Figure 3. Alternatively, following the lead of H17, we tested for steady period change by fitting a quadratic to these times. The resultant best-fit curve is also shown in Figure 3, and corresponds to

$$\text{Mid-eclipse} = \text{BJD(TDB)}\ 2{,}437{,}547.72884(3) + 0.0566878480(1)\ E - 2.9(4) \times 10^{-15}\ E^2. \qquad (3)$$

The quadratic term corresponds to $dP/dt = 8.5 \times 10^{-14}$, or a timescale for period decrease $P/(dP/dt) = 1.8(9) \times 10^9$ years. This is nominally 3 times slower than the period decrease derived by H17. But more importantly, both the linear and quadratic fits are obviously poor descriptions of the data — with the apparent period change (slope of the O–C) in recent years being opposite to that of earlier years. The visual impression suggests something more like a quasiperiodic wiggle.

We were disappointed in this result. The analysis by H17 suggested that it might be possible to learn the direction of evolution through the observed change in orbital period. That would certainly be a welcome empirical constraint on the many theoretical studies of the evolution of short-period CVs. But the above discussion shows that this particular reward cannot yet be reaped from a study of the O–C residuals. Addition of more timings, later timings, and correction for the time standard (leap seconds) have vitiated the H17 result.

## 4  OR MAYBE PHASE CHANGE?

In a (probably quixotic) search for any periodic signature in the residuals from the linear fit, we calculated the power spectrum of these points.  The result is seen in the upper frame of Figure 5, where the candidate peaks are labeled with their frequencies in cycles/year.  Among the labeled peaks, one occurs at a frequency corresponding to the inter-outburst period (22.6 years), and several others are consistent with harmonics of that frequency.  That would be consistent with a signal which is significantly non-sinusoidal.  So it is worth considering the hypothesis that the main source of long-term variance in Figure 3 is "place in the outburst cycle".  The lower frame of Figure 5 shows a fold of the O-C curve on the 22.6 year cycle[25], where zero phase is the sudden rise to outburst on 1 December 1978.

While far from inspiring, and still afflicted by the seasonal noise suggested in Figure 4, this seems to be worth exploring.  In particular, could this far-from-inspiring-and-not-provably-periodic drift be merely a phase change, not directly signifying the actual orbit?

The assumption that mid-eclipse is a stable marker of true dynamical phase is certainly open to question.  Radial-velocity evidence shows that the eclipse marks the transit of the secondary across the line of sight to the "hot spot", presumably where the mass-transfer stream strikes the periphery of the accretion disk (Krzeminski & Smak 1971, and all later studies).  That transit time would be affected by the size and shape of the disk, and the shape of the hot spot.  Physics offers no guarantees that such parameters are stable in a cataclysmic variable; and indeed, these parameters are demonstrably unstable in the actual outburst state of WZ Sge (Patterson et al. 2002).

The main feature of the O-C plot in Figure 5 is the sudden drop by ~0.0025 cycles at outburst phase zero.  Can this be plausibly interpreted in terms of the outburst physics?

Yes, probably it can.  With the approximation that the hot-spot is a point at the exact periphery of a circular disk, where the ballistic stream strikes the disk, the mid-eclipse time reveals the size (radius) of the disk.  A very large disk implies eclipse near dynamical phase zero (inferior conjunction of the donor star), while a somewhat smaller disk produces slightly later eclipses.[26]  Figure 5 of RNP illustrates the geometry. Therefore the sudden (within one year) drop in O–C could mean that the disk has expanded ... and then slowly contracts over the following 22.6 years. That is not inconsistent with our current understanding of dwarf-nova physics (in particular, the expansion of the disk during outburst, as it transitions to a high-viscosity state).

---

[25] WZ Sge has had 4 outbursts during 1913–2001, so the cycle is usually (and properly) considered as ~30 years.  But the two outbursts (1978 and 2001) in this data occurred almost exactly in the middle of our interval of coverage (1961–2017), so 22.6 years seems like the right number.

[26] This is only true for disks which fill an appreciable fraction of their Roche lobes.  A very tiny disk would have a hot-spot very close to the white dwarf (WD), and eclipse would then occur very close to dynamical phase zero.  The measured radial velocities at the hot spot (~700 km/s) suggest a fairly large disk.

There is another advantage to this kind of explanation. A true change in the orbital period of a binary star requires moving around large quantities of energy and angular momentum. Marsh and Pringle (1990) discuss this problem in detail. In this low-luminosity and very tight binary with a puny (< 0.08 $M_\odot$) donor star, such quantities are hard to find — especially if they have to be moved around *cyclically*. Changes in orbital phase present no such difficulty, especially if they are linked to a cycle known by other means (the historical light curve) to be present in the star.

## 5 LESSONS FROM OTHER STARS

But there are similar studies (O–C analyses) of other CVs and related stars with sharp-walled eclipses, which are demonstrably eclipses of the white dwarf (WD) — and therefore are very likely markers of true dynamical phase. What do their O–C diagrams look like?

Pretty much like that of WZ Sge. Some evidence for orbital period change has been presented for the short-period dwarf novae OY Car (Han et al. 2015), HT Cas (Borges et al. 2008), V2051 Oph (Baptista et al. 2003, Qian et al. 2015), DV UMa (Han et al. 2017b) and Z Cha (Baptista et al. 2002, Dai ei al. 2009). All these stars appear to show O–C wiggles on timescales of 10-30 years, and the first three also had a claim of orbital-period decrease. Since the decades-long timescale of the wiggles is similar to the time span of the observations (15–45 years), the two effects — cyclic and monotonic — are still not clearly established and separated.

Also, these 5 stars show a blend of WD and hot-spot eclipse, and these components are not simple to decouple, since they overlap in time. Thus they are not altogether immune from the vagaries introduced by the hot spot. And since they're not quite immune, the question arises: are these true period changes, or merely "phase changes"?

Perhaps an instructive lesson here can be learned from the O–C diagrams of post-common-envelope binaries (PCEBs[27]), which are "clean" systems — with simple WD eclipses, uncontaminated by any accretion-disk effects. The light curves are easy to analyze (see Figures 2–5 of Parsons et al. 2010), and the dispersion of seasonal eclipse timings in PCEBs is sometimes as low as ~1 s, compared to ~5–10 s for actual mass-transfer systems (CVs). What do their O–C diagrams show?

This has been the subject of several lucid discussions (Beuermann et al. 2010, Parsons et al. 2010, Zorotovic & Schreiber 2013, Marsh et al. 2014). Most PCEBs show alternating

---
[27] They could also be considered "pre-cataclysmic eclipsing binaries", since their orbital periods and separations are small enough to drive the WD and red dwarf into Roche-lobe contact on short timescales, via GR.

O–C wiggles of a similar size (amplitudes ~20 s, periods or quasiperiods 5–20 years), and these are most commonly interpreted as light-travel-time effects from distant orbiting planets. But many of these planetary interpretations have not survived, either because the claimed orbits are not stable, or because subsequent eclipse timings deviate markedly from predictions. Marsh et al. (2014, section 4.2) summarizes the present situation well.[28]

But the O-C wiggles are still there in PCEBs, despite the absence of accretion-disk contamination, and despite failures of human ingenuity to explain them. And the size and time-scale of the wiggles are similar to those found in CVs (including WZ Sge). So, following a well-known principle of scientific economy (*pluralitas non est ponenda sine necessitate*, Ockham 1320), one might turn to the one ingredient which is certainly common to all these binaries: the presence of a low-mass convective star, in rapid rotation enforced by the Roche-lobe synchronism. Such stars may well have "activity cycles". This is thought be the origin of some (many?) O–C variations in Algol binaries, and conjectured for CVs as well (Warner 1988; Applegate 1992; Richman, Applegate, & Patterson 1994). As detailed by Richman et al., quasiperiodic and decades-long variations in brightness and eclipse times are very commonly seen in cataclysmic variables of all types.

## 6 SUMMARY

1. We report 44 new timings of mid-eclipse, which extend the baseline, fill gaps during years of sparse reporting, and improve timing accuracy. Outside the post-eruption years (1978–9 and 2001), the "quiescent" light curves and eclipse profiles have not changed significantly.

2. Analysis of timings reveals a best-fit period of 0.05668784695(8) d, but the fit is obviously poor. The phase appears to slowly wander by ~0.002 cycles (9 s), on a timescale of decades. This could possibly be related to the ~23 year interval between eruptions. Regardless of its true origin, this apparent back-and-forth phase wander prevents a robust measurement of long-term secular ("evolutionary") period change, such as that discussed by H17. This is especially unfortunate, because a secure measure of monotonic orbital-period change is the only sure test of "period bounce" — which has become the Holy Grail of CV evolution.

3. Even the *search* for that period change in WZ Sge has poor prospects, since the eclipsed object appears to be the hot spot, not the WD.

4. Similar (nonperiodic, on timescales of 5–30 years) wanderings in O–C have been reported in many other CVs, even when the central eclipsed object is definitely the WD. Indeed, some (apparently) nonperiodic O–C wanderings even occur in PCEBs, which have no accretion disk and therefore reveal precisely the time of dynamical conjunction. The commonality of this wander suggests an origin in the one ingredient present in all three classes: the low-mass secondary stars.

---

[28] See also Zorotovic & Schreiber (2013), especially their Section 5.4. These two references are required reading on the subject.

This research was supported in part by NSF grants AST12–11129 and AST16–15456 to Columbia University.

**FIGURE CAPTIONS**

Figure 1.  *Lower frame:*  a typical light curve, spliced from two CBA stations.  The splice is made by normalizing the out-of-eclipse magnitudes to zero (with additive constants).  *Upper frame:*  the mean orbital light curve in 2017, showing the sharp but shallow eclipse.

Figure 2.  The mean orbital light curve near eclipse, at 5-s time resolution, during a 15-orbit average of the 2017 small-telescope data (obtained with ~30 s sampling).  At each integration's mid-point, integrative sampling is equivalent to instantaneous sampling, except that in theory, very rapid changes are not quite resolved.  For purposes of this study, all the relevant features are rendered exactly: mid-ingress, mid-egress, and the slight asymmetry between ingress and egress.  Averaging over orbits also helps subdue flickering.  The eclipse width (mid-ingress to mid-egress) here is 172±10 s, consistent with the estimate of 160±9 s in the best published data (Robinson et al. 1978).

Figure 3.  O–C diagram of the mid-eclipse timings, relative to Eq. (1).  The line and parabola indicate least-squares fits to the timings, and correspond to Eqs. (2) and (3).  Arrows indicate times of the 1978 and 2001 outbursts (usually called "superoutbursts" in the dwarf-nova literature, but we shorten the term here, since normal outbursts have never been seen in the star's vast observational record).

Figure 4.  Expanded view of the 1976 and 1977 portions of Figure 3.  All these timings are from RNP — the best published data in the historical record.  The between-years shift of 12±3 s illustrates "seasonal noise" in the timings.  This presumably exists every year, but is not obvious in less extensive or more noisy data sets.

Figure 5.  *Upper frame:*  power spectrum of the residuals from Eq. (2).  The frequencies of several candidate peaks are identified (in cycles/year).  These are close to harmonics of the outburst frequency (1/22.6 years) in this interval.  *Lower frame:*  fold of the residuals on the 22.6 year period.

# TABLE 1

**Timings of Mid-Eclipse**

*Mid-Eclipse*
[BJD(TDB) − 2,400,000+]

| | | | | | |
|---|---|---|---|---|---|
| 37547.72863 | 37554.75812 | 37613.65698 | 37816.88272 | 37880.76986 | 37912.79846 |
| 37937.79801 | 37938.64816 | 38580.58149 | 38623.43732 | 38624.57117 | 39294.96176 |
| 39296.94585 | 39325.79990 | 39327.84055 | 39704.81483 | 39705.77849 | 39707.76255 |
| 40416.53067 | 40417.55122 | 40418.51497 | 41208.68675 | 41208.74343 | 41210.61399 |
| 42930.80684 | 42930.86352 | 42931.82719 | 42931.88380 | 42931.94057 | 43011.70027 |
| 43011.75699 | 43011.81376 | 43013.68449 | 43013.74115 | 43013.79780 | 43047.64047 |
| 43047.69701 | 43311.80597 | 43311.86274 | 43311.91945 | 43319.91214 | 43321.89632 |
| 43348.82311 | 43364.75244 | 43364.80904 | 43364.86592 | 43368.77717 | 43368.83395 |
| 43368.89054 | 43395.70389 | 43395.76061 | 43395.81723 | 43396.61096 | 43396.66750 |
| 43396.72433 | 43396.78099 | 43396.83779 | 43400.63572 | 43400.69243 | 43401.65615 |
| 43401.71280 | 43635.83351 | 43635.89040 | 43640.82221 | 43659.86939 | 43714.68643 |
| 43714.74295 | 43714.91319 | 43717.69096 | 43721.88582 | 43722.84953 | 43722.90619 |
| 43727.89470 | 43729.82214 | 43731.69268 | 43731.74942 | 43731.86282 | 43747.84887 |
| 43748.75593 | 43760.83030 | 43761.62406 | 45143.84375 | 45143.90039 | 45286.58370 |
| 47391.68699 | 47391.74353 | 47392.65069 | 47392.70704 | 47392.76400 | 47392.82070 |
| 48484.68533 | 48485.64894 | 48485.70571 | 48485.76223 | 49898.70710 | 49904.6592 |
| 49904.7160 | 49904.7725 | 49904.8294 | 49905.6796 | 49906.7001 | 49906.7566 |
| 50275.79437 | 50276.75822 | 50281.63335 | 50281.69014 | 50281.74692 | 50281.80367 |
| 50282.71055 | 50282.76720 | 50282.82390 | 50283.73087 | 50283.78734 | 50287.69895 |
| 50287.75569 | 50303.57160 | 50306.57611 | 50310.60091 | 50310.71441 | 50311.67790 |
| 50311.7347 | 50312.64160 | 52382.99509 | 52542.79802 | 52543.64850 | 52543.76190 |
| 52774.93479 | 52775.95520 | 52776.97561 | 52778.90291 | 52779.92326 | 54777.99963 |
| 54781.00405 | 54796.99031 | 54799.99471 | 54801.01513 | 55046.87021 | 55064.72702 |
| 55423.78779 | 57171.36074 | 57254.57856 | 57257.41296 | 57536.37366 | 57705.07666 |
| 57951.61233 | 57956.03409 | 57968.05195 | 57978.59575 | 57981.60032 | 57983.47094 |
| 57985.68184 | 57986.41859 | 57989.59330 | 57990.72704 | 57992.82445 | 57993.73150 |
| 57996.73605 | 58004.72886 | 58008.69706 | 58022.58567 | 58027.57428 | 58028.65116 |
| 58029.38833 | 58030.40868 | 58038.68490 | 58042.65300 | 58044.58052 | |

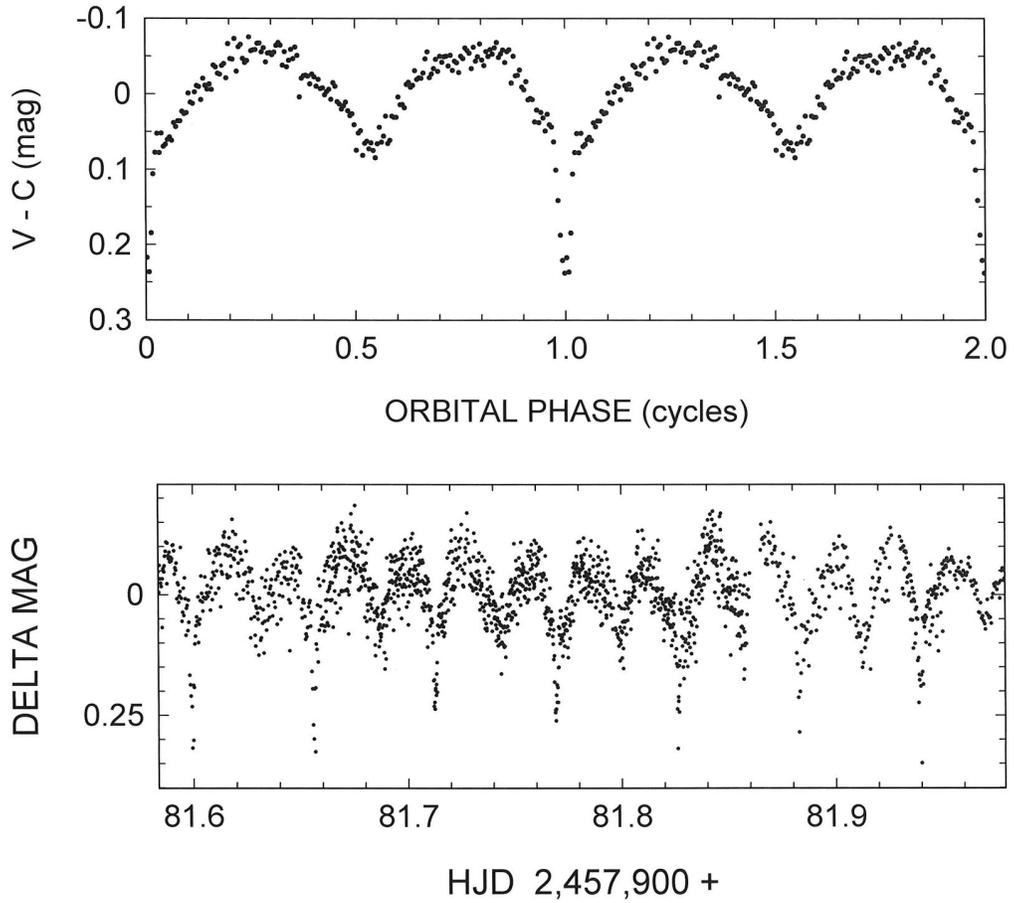

Figure 1. *Lower frame:* a typical light curve, spliced from two CBA stations. The splice is made by normalizing the out-of-eclipse magnitudes to zero (with additive constants). *Upper frame:* the mean orbital light curve in 2017, showing the sharp but shallow eclipse.

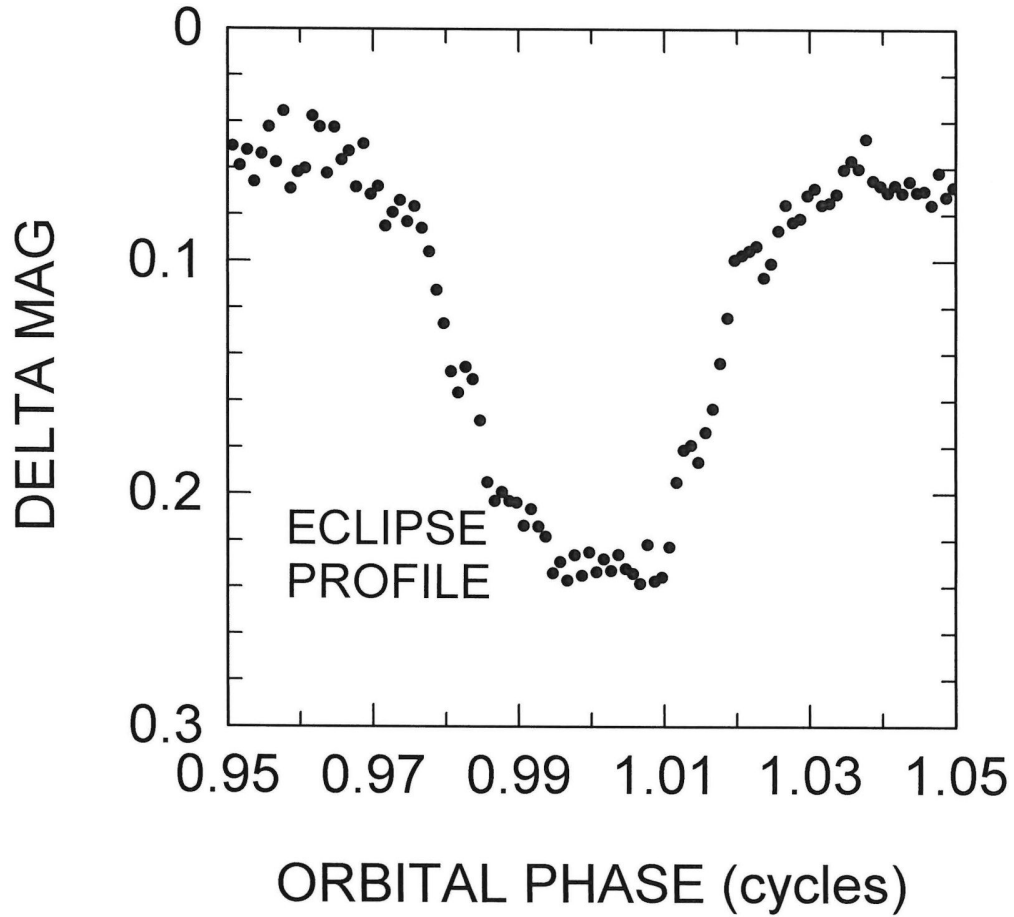

Figure 2. The mean orbital light curve near eclipse, at 5-s time resolution, during a 15-orbit average of the 2017 small-telescope data (obtained with ~30 s sampling). At each integration's mid-point, integrative sampling is equivalent to instantaneous sampling, except that in theory, very rapid changes are not quite resolved. For purposes of this study, all the relevant features are rendered exactly: mid-ingress, mid-egress, and the slight asymmetry between ingress and egress. Averaging over orbits also helps subdue flickering. The eclipse width (mid-ingress to mid-egress) here is 172±10 s, consistent with the estimate of 160±9 s in the best published data (Robinson et al. 1978).

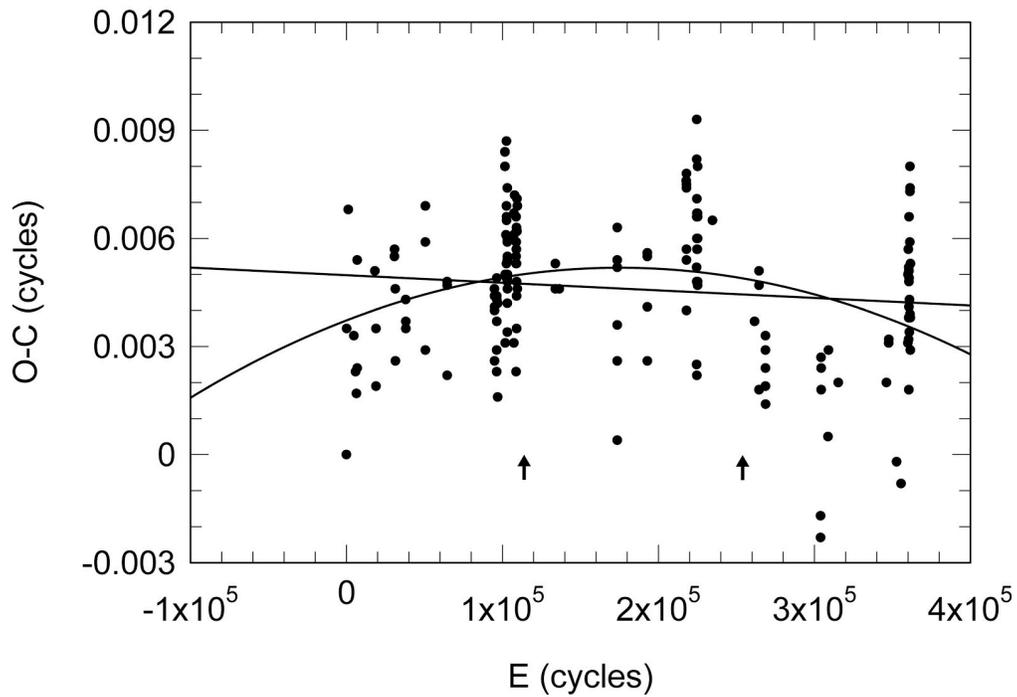

Figure 3. O–C diagram of the mid-eclipse timings, relative to Eq. (1). The line and parabola indicate least-squares fits to the timings, and correspond to Eqs. (2) and (3). Arrows indicate times of the 1978 and 2001 outbursts (usually called "superoutbursts" in the dwarf-nova literature, but we shorten the term here, since normal outbursts have never been seen in the star's vast observational record).

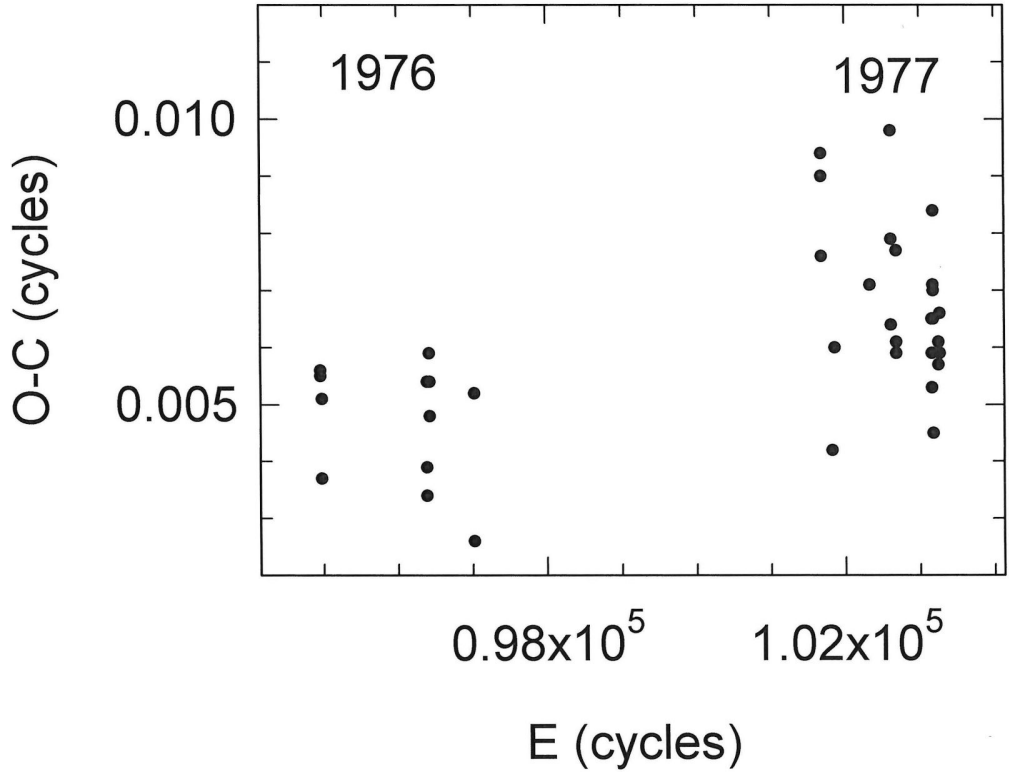

Figure 4. Expanded view of the 1976 and 1977 portions of Figure 3. All these timings are from RNP — the best published data in the historical record. The between-years shift of 12±3 s illustrates "seasonal noise" in the timings. This presumably exists every year, but is not obvious in less extensive or more noisy data sets.

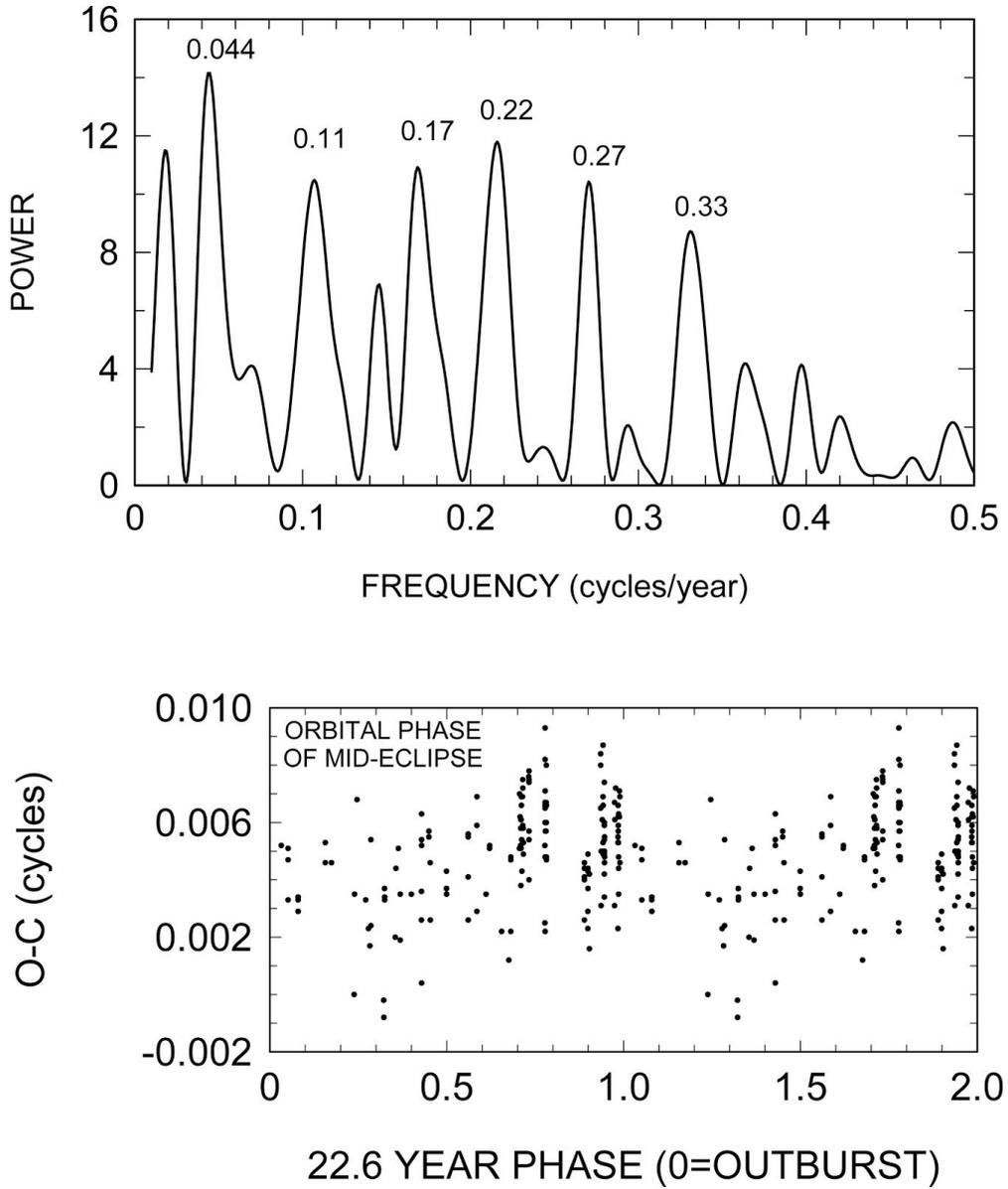

Figure 5. *Upper frame:* power spectrum of the residuals from Eq. (2). The frequencies of several candidate peaks are identified (in cycles/year). These are close to harmonics of the outburst frequency (1/22.6 years) in this interval. *Lower frame:* fold of the residuals on the 22.6 year period.